# Active correlations technique: Schmidt equation with partially free order

Yu.S.Tsyganov

1. **Introduction**

   $^{48}$Ca - induced fusion reactions on heavy actinide targets have been investigated since 1999 at the FLNR, Dubna [1]. Results on the discovery of many isotopes and new superheavy elements (SHE) Z=114,115,116,118 convincingly were reported since 2004 and presented in a review [2]. Recently, the discovery of $^{283}$112 was confirmed [3,4]. Production cross-sections observed at FLNR reached a level 5 pb for elements Z=114 and Z=116. The 1pb level was reported for the formation of Z=112 and Z=118. Also the odd elements Z=115 and Z=113 were produced in the range (1-3) pb. The mentioned orders of measured cross-sections assume that not only experimental apparatuses, especially the detection systems are operate well during the long term experiments, but some statistical models for results interpretations are of appropriate quality. All these results were obtained at the Dubna Gas Filled Recoil Separator (DGFRS) – the mostly advanced facility of FLNR [5].

2. **Statistics models for rare detected sequences**

   In nuclear physics, especially in the experiments aimed to the discovery SHE (or/and isotopes), the technique of delayed coincidences is widely used for detecting time-energy-position correlations between signals of different groups.

   There are two aspects in establishing the significance for the existing of true correlation:

   • Consideration of the possibility that the random background of uncorrelated events could simulate a correlation is required.

   • Estimation of the compatibility of the parameters of the observed events with known properties of some numbers in the considered event chain should be under consideration too.

   It was K.H. Schmidt who first recognized the mentioned problems and epitomized a compact theoretical approach for numerical consideration [6].

   Note that additionally to this approach, another theoretical model was formulated in [7,8 ].

   These theories are known as **_LDSC_** (**_L_**inked **_D_**ecay **_S_**ignal **_C_**ombinations) and **_BSC_** models, (**_B_**ackground **_S_**ignal **_C_**ombinations) respectively.

   Having mentioned to these approaches as classical, one should take into account the existence of a different approach to the problem basing on some Monte Carlo calculations [9].

The goal of the present paper is to consider application of an approach [5] relatively the experimental method of "active correlations" which was extensively used to suppress background signals in heavy ion induced nuclear reactions [10-14]. An approach, describing in the Ref. [9] is outside the scope of the present paper.

3. Correlation analysis: Schmidt equation for correlation chain with partially free order

The a priori knowledge of the order of the events in a possible true event chain may be limited. In [6] ( sect.3.2) one case which is characterized by the condition that possible decay sequences are known to start with the events $E_1$ of the group 1 is considered. The events of the other event groups ($E_2$ to $E_K$) may appear in any order, but at least one event $E_i$ must appear within the time limit $\Delta t_{1,i}$. The equation for numbers of random events $n_b$ was obtained in the form:

$$n_b = \lambda_1 T \prod_{i=1}^{K-1} \left\{ \int_0^{\Delta t_{1,i+1}} (dp_{1,i+1}/dt) dt \right\}, \qquad (1)$$

where T is an effective time of the experiment ( see [6,9]), $\lambda_i$ – rate of events of i-type, K – number of chains in the multi event, $dp_{1,i}/dt$ is the probability density that an event $E_1$ is followed by an event $E_i$ after the time distance t. For the case of the condition ($\lambda_i + \lambda_1$)$\Delta t_{1,i} \ll 1$ is true the above written equation can be simplified as

$$n_b \approx T \prod_{i=1}^{K} \lambda_i \prod_{i=2}^{K} \Delta t_{1,i}. \qquad (2)$$

4. The DGFRS Detection Module

The Dubna Gas Filled Recoil Separator is the most efficient facility in use in the field of synthesis of superheavy elements of Flerov Laboratory of Nuclear Reactions of JINR [5]. Their separation characteristics are based on the ion optical properties of the gas-filled magnetic dipole. For the synthesis and study of heavy nuclides, the complete fusion reactions of target nuclei with heavy bombarding projectiles are used. The resulting excited compound nuclei can de-excite by evaporation of a few neutrons, while retaining the total number of protons. Recoil separators are widely used to transport evaporation residues from the target to the detection system, simultaneously suppressing the background products of other reactions, the incident ion beam, and scattered target nuclei. Of course, the detection module should not contribute to the significant decreasing of the mentioned parameter. The detection system of the DGFRS consists of silicon 12 strip position sensitive detector to measure recoils energy end their forthcoming alpha (spontaneous fission) decays, and low pressure pentane filled time-of-flight module to detect charged particles incoming from cyclotron. The detection module of the

DGFRS is shown in the Fig.1. Silicon "veto" detector, consisting of three separate silicon chip, is placed behind the focal plane one to detect charged long path particles passing through the main position sensitive detector and creating no any signal in the gaseous TOF module.

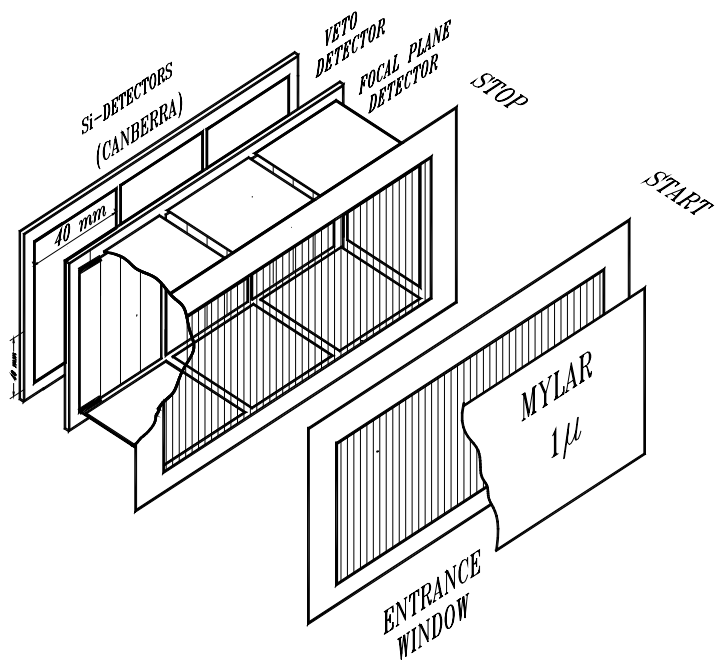

**Fig.1** The detection module of the DGFRS (Schematics).

Namely with the described detector it has became possible to establish in the real SHE experiment the genetic links between the different chains of multi - chain event.

### 5. Method of "active correlations" for beam associated backgrounds suppression

Usually, to reach high total SHE experiment efficiency, one use extremely high (n*$10^{12}$ to $10^{13}$ pps, n > 1) heavy ion beam intensities. It means, that not only irradiated target, sometimes (frequently) made on highly radioactive actinide material, should not be destroyed during long term experiment, but the in-flight recoil separator and its detection system should provide backgrounds suppression in order to extract one-two events from the whole data flow. Typically, the DGFRS provides suppression of the beam-like and target-like backgrounds by the factors of ~$10^{15}$-$10^{17}$ and $10^4$-5*$10^4$, respectively. Nevertheless, under real circumstances, total counting rate above approximately one MeV threshold is about tens to one-three hundreds events per second. Therefore, during, for example one month of irradiation about 30*$10^5$*100 =3e+08 multi - parameter events are written to the hard disk during a typical SHE experiment.

To avoid a scenario that result of the SHE experiment (one-two-three decay chains per month) can be represented as a set of random signals real-time search technique to suppress the probability for detected event to be a random has been designed and successfully applied.

Note, that in the reactions with $^{48}$Ca as a projectile, the efficiency of SHE recoiling products detection both by silicon and TOF detector is close to 100%. Namely recoil-first (second) correlated alpha decay signal was used as a triggering signal to switch off the cyclotron beam for a definite (seconds-minutes) and, therefore, detection of forthcoming alpha decays were in fact "background-free". The basic idea to apply such detection mode is to transform the main data flow to the discrete form [10-14]. To demonstrate this technique successful application in the Fig.2 the spectrum of α-decays detected in $^{237}$Np+$^{48}$Ca complete fusion reaction is shown [15].

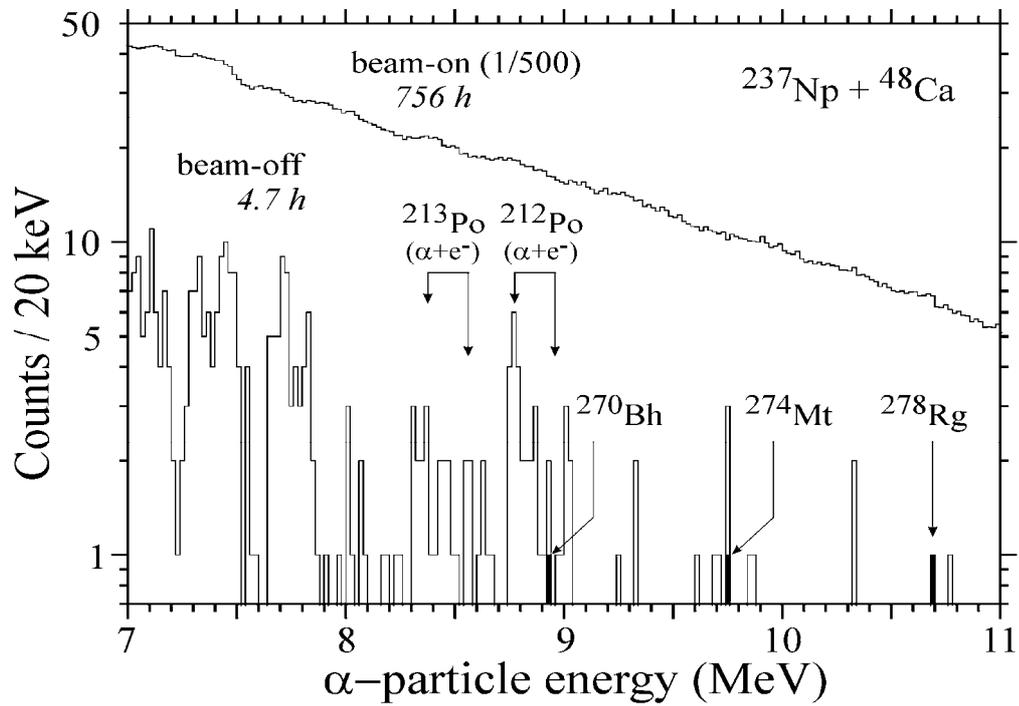

**Fig.2** Beam-OFF spectrum of alpha decays was measured in $^{237}$Np+$^{48}$Ca complete fusion reaction [15].
Energy ranges: for EVR (7, 18) and for alpha particles (9.9, 11.4) MeV. Correlation time was 1.5 s for y-position window 2.8 mm. Beam-Off intervals was of 10 s.

Being considering in the described above process a definite order correlated pair recoil-alpha $E_1 \rightarrow E_2$ as a starter signal $\hat{E}_1 \equiv (E_1 \cap E_2)$ for forthcoming sequences of "α"-decays and following to the philosophy of [6] one can rewrite the equation (1) for the given case in the form of:

$$n_b = \hat{\lambda}_1 T \prod_{i=2}^{K-1} \{ \int_0^{\Delta t_{2,i+1}} (dp_{2,i+1}/dt)dt \} \qquad (3).$$

Here the parameter $\hat{\lambda}_1$ denotes not any single signal rate per pixel, but a rate of correlations/pauses generating by the detection system during a long term experiment. Therefore, if $N_{STOP}$ is a total number of pauses measured in an experiment, to a first approximation one can consider $N_{STOP} = \psi \hat{\lambda}_1 T$ and:

$$n_b = \psi \cdot N_{STOP} \prod_{i=2}^{K-1} \{ \int_0^{\Delta t_{2,i+1}} (dp_{2,i+1}/dt)dt \} \qquad (4).$$

The simplified formula (2) is rewritten as:

$$n_b \approx \psi \cdot N_{STOP} \overline{\lambda}^{K-2} \prod_{i=3}^{K} \Delta t_{2,i} \qquad (5),$$

where $\overline{\lambda}$ - a mean counting rate value for alpha – decay signals measured in beam-OFF pauses by the focal plane detector.

If to take into account more common case of detecting of α-particles by side detector and with finishing spontaneous fission signal, one should rewrite (5) in the form of (5'):

$$n_b \approx \psi \cdot N_{STOP} \overline{\lambda}^{K-3-m} \cdot \overline{\lambda}_{ESC}^{m} \cdot \lambda_{FF} \prod_{i=3}^{K} \Delta t_{2,i} \qquad (5').$$

In this equation $\overline{\lambda}_{ESC}$ - mean rate per detector of escaping α-particles, $\lambda_{FF}$ – rate of fission fragment signals imitators per pixel, m- number of detected out off beam escaping α-particles. Of course, in (5) – (5') it is assumed, like in [6] that $(\lambda_2 + \lambda_i) \cdot \Delta t_i \ll 1$.

Parameter of ψ is denotes an effective time part[1] $\psi \approx \dfrac{2t_{EVR-\alpha}}{t_{PRS}}$, where t EVR-α is the measured recoil-alpha correlation time and $t_{PRS}$ is the pre-setting time parameter ( t $_{PRS}$ > t $_{EVR-\alpha}$, or even t $_{PRS}$ >> t $_{EVR-\alpha}$) for beam stopping process. Factor two for an optimal interval is explained in [8]. Or, in the case of non-uniform distribution: $\psi = \int_0^{\frac{m+1}{m}t_{EVR-\alpha_m}} (\dfrac{dN_{STOP}}{dt})dt$, where m is a number of α-particle signals creating beam-stop process and $\hat{E}_1 \equiv (E_1 \cap E_2 \cap E_3 ... \cap E_{m+1})$ (see Ref. [6]).

Note that in [16] nearly the same conclusion was drawn by using BSC philosophy and on separate elementary events spaces representation for groups "beam ON" and "beam OFF". It is not excluded that the higher order correlated event may be considered by a similar way as an event "starter".

---

[1] For a flat distribution

## 6. Summary

Schmidt equation with partially free order was used to obtain estimate formulae for the case of application of an active correlations technique. These relations one can use to estimate of statistical significance of rare decay events measured in experiments aimed to the synthesis of SHE. Additionally, the present formulae may use to establish the limitations for the method application in different heavy ion-induced nuclear reactions. Author plans to extend the philosophy presented in the present paper in a nearest future. Also, the modification of the whole active correlation technique is in progress now. Author also plans to built an experimental setup to check the basic consequences of statistical approaches which required no beam time for detection of α-α correlated sequences [17].

From a more common viewpoint the present paper denotes a first step to create a rare statistics theory of strongly non steady state stochastic processes which describes applications of active data acquisition system in the long term experiments.

This work is supported in part by **RFBR** grant №07-02-00029.
Author is indebted to Drs. A.Polyakov, K.Subotic, V.Utyonkov for their support of the preset paper.

**Supplement 1**

**Partially definite order for begin and end of a multi-event.**

Definitely, one can speak about any definite order if:
a) EVR is a first signal of multi chain event, and first alpha particle signal provides correlation, which stops the target irradiation process;
b) Reasonable, for a last alpha-SF chain, because of SF-α chain is unreasonable.

The middle part of the entire event one could consider as with definite order in the case of to any signal one can attribute a definite isotope/peak. It is reasonable for a not pure statistics and if the measured signals are resolved to each other. Therefore, in the field of unknown nuclides and very rare statistics and if energy intervals is high enough, up to one – two MeV's, let us consider these group in a free order.

So, in that case:

$$n_b = \psi \cdot N_{STOP}(1 - e^{-\lambda_F \Delta t_{\alpha-SF}}) \prod_{i=2}^{K}(1 - e^{-\lambda_i t_{0,i}}) \prod_{j \neq i}^{K} e^{-\lambda_j t_{0,j}}$$

$$n_b \approx \psi \cdot N_{STOP} \cdot (1 - e^{\lambda_F t_{\alpha_N-SF}}) \cdot \overline{\lambda}^k \cdot \overline{\lambda}_S^{N-k-1} \prod_{i=2}^{N} \Delta t_{1,i} \approx \psi \cdot N_{STOP} \overline{\lambda}^k \overline{\lambda}_S^{N-k} \lambda_{SF} \Delta t_{\alpha_N-SF} \cdot \prod_{i=2}^{N} \Delta t_{1,i}$$

Subscript "S" is considered as "side detector" (mean value), N - number of α-particle signals in a multi chain event starting from 1. It is not excluded, that the measured time interval value is not optimal, and factor (n+1)/n should be taken into account [8] as a normal maximum likelihood estimate of the interval size, where n – number signals except for a "starter" (EVR).

**Supplement 2**

**Comparison table for LDSC and BSC methods**

| LDSC | BSC |
|---|---|
| A priori information about multi chain event | No a priori information |
| Time interval values between single signals are actual | Only a total time interval is actual |
| No invariance (part. free order equation) | Invariant with t → - t |
| Probability estimate is stable with $\Delta t_{\alpha\text{-SF}} \to \infty$ | Non stable |
| Sensitivity to definite order | No sensitivity |
| Approach is more differential | More integral method |
| Nearly the same | Stochastic Poisson's process is used for a data representation |
| The value of effective time is introduced in the form T = f·t, where t – duration of experiment, f – number of effective detector pixels (see e.g. Ref.[9]), taking into account strip number and position resolution | Only one pixel consideration for an expectation value is used, but one may correct $N_m$ by using **LDSC** effective time in the manner of Ref. [9]. So, value of $N_m$ = 0.0004 reported in [8] one should replaced by $\widetilde{N}_m \approx f \cdot N_m \approx 200 \cdot N_m \approx 200 \cdot 0.0004 \approx 10^{-1}$ for the entire detector |

**References**

[1]  Armbruster P. // Eur. Phys. J. A 37, (2008) 159 -167

[2]  Oganessian Yu.Ts. et al. // J. Phys. G. Nucl. Part. Phys. 34, (2007) R.165

[3]  Eichler R. et al. // Nature 83, (2007) 487

[4]  Hofmann S. et al.// Eur. Phys. J., A 32 (2007) 251

[5]  Subotic K. et al.// Nucl. Instrum. and Meth. In Phys. Res. A . 2002. V.481 p.71-80

[6]  Shmidt K.H. et al.// Z. Phys. A. 316 (1984)19-26

[7]  Zlokazov V.B. // Eur. Phys. J. A 14 (2002) 147

[8]  Zlokazov V.B.// Yad. Phys. 2003, v.66b №9, (2003) pp.1714-1718

[9]  Stoyer N. et al. // Nucl. Instrum. and Meth. In Phys. Res. A455 (2000) 433

[10] Tsyganov Yu.S.// Lett to ECHAYA, v.4 №4 (140) (2007) p.608-613

[11] Tsyganov Yu.S.// Lett to ECHAYA, v.6, №1(150) (2009) p.96-102

[12]  Tsyganov Yu.S. et al.// Nucl. Instrum. and Meth. In Phys. Res. A 513 (2003) 413-416

[13] Tsyganov Yu.S. et al.// Nucl. Instrum. and Meth. In Phys. Res. A 525 (2004) 213-216

[14] Tsyganov Yu.S. //Nucl. Instrum. and Meth. In Phys. Res. A 573 (2007) 161-164

[15] Oganessian Yu.Ts. et al.// Phys. Rev. C 76, (2007) 011601 (R)

[16] Tsyganov Yu.S. // JINR communication P13-2008-92

[17] Nadderd L., Tsyganov Yu.S, Subotic K. et al. // in Proc. Of XXI Int. Symp. Nuclear Electronics & Computing, NEC'2007. Varna, Bulgaria, 10-17 Sept. 2007.
ISBN 5-9530-0171-1. Dubna 2008. pp.362-366